\documentclass[11pt,a4paper]{article}
\usepackage{graphicx,amssymb,bm}

\begin{document}

\title{ON A POWER SERIES INVOLVING CLASSICAL ORTHOGONAL POLYNOMIALS} 
\author{Paulina Marian\,$^{1,2}$,
Tudor A. Marian\,$^1$\\
$^1$ Centre for Advanced Quantum Physics,\\ 
Department of Physics, University of Bucharest, \\
P. O. Box MG-11, R-077125 Bucharest-M\u{a}gurele, Romania\\
E-mail: tudor.marian@g.unibuc.ro\\
$^2$  Department of Chemistry, University of Bucharest, \\
Boulevard Regina Elisabeta 4-12, R-030018 Bucharest, Romania\\
E-mail: paulina.marian@g.unibuc.ro}
\maketitle
\begin{center}
(Received \today )
\end{center}

\begin{abstract}

We investigate a class of power series occurring in some problems in
quantum optics. Their coefficients are either Gegenbauer or Laguerre 
polynomials multiplied by binomial coefficients. Although their sums 
have been known for a long time, we employ here a different method 
to recover them as higher-order derivatives of the generating function
of the given orthogonal polynomials. The key point in our proof consists 
in exploiting a specific functional equation satisfied by the generating 
function in conjunction with Cauchy's integral formula for the derivatives 
of a holomorphic function. Special or limiting cases of Gegenbauer 
polynomials include the Legendre and Chebyshev polynomials. The series 
of Hermite polynomials is treated in a straightforward way, as well as 
an asymptotic case of either the Gegenbauer or the Laguerre series. 
Further, we have succeeded in evaluating the sum of a similar power series 
which is a higher-order derivative of Mehler's generating function. 
As a prerequisite, we have used a convenient factorization of the latter 
that enabled us to employ a particular Laguerre expansion. 
Mehler's summation formula is then applied in quantum mechanics 
in order to retrieve the propagator of a linear harmonic oscillator. 
\end{abstract}


{\centering\section{INTRODUCTION}}

The problem of evaluating the $l$th-order correlation function
$\langle({\hat a}^\dag)^{l}{\hat a}^{l}\rangle$ for a one-mode squeezed 
thermal state of the quantum electromagnetic field by employing 
the photon-number distribution led one of us \cite{PM} to consider 
the following power series involving Legendre polynomials:
\begin{equation}
G_{l}^{(\frac{1}{2})}(t,w):=\sum_{n=l}^{\infty}\left(\begin{array}{c}n\\l
\end{array}\right)t^{n}P_{n}(w), \qquad (w:=\cos\theta,
\;\;\; |t|<1, \;\;\; l=0,1,2,3,...). \label{Legser}
\end{equation}
In eq.~(\ref{Legser}), $\left(\begin{array}{c}n\\l\end{array}\right)$ 
are binomial coefficients. For subsequent convenience, we introduce 
the square root of a complex variable as follows:
\begin{equation}
q:=(1-2wt+t^2)^{\frac{1}{2}}, \qquad  (t=0 \longrightarrow q=1).
\label{var}
\end{equation}
The sum of any series of ascending powers \ (\ref{Legser}) is the product 
of three factors: the generating function of the Legendre polynomials, 
the $l$th power of a specific variable, and the Legendre polynomial 
of degree $l$ of another specific variable:
\begin{equation}
G_{l}^{(\frac{1}{2})}(t,w)=\frac{1}{q}
\left(\frac{t}{q}\right)^l
P_{l}\left(\frac{w-t}{q}\right), \qquad (l=0,1,2,3,...).  
\label{Legsum}
\end{equation}
Note that the series expansion\ (\ref{Legser}) reduces for $l=0$ 
to the power series of the generating function of the Legendre polynomials.

A similar result has been found later for the Laguerre polynomials 
\cite{PTB}. We obtained a formula which is useful when the general 
solution of the quantum optical master equation describing dissipation
is applied to a single-mode displaced squeezed thermal state \cite{PTA}.

When we performed the above-mentioned summations we were not aware 
of the fact that the corresponding formulae and the similar ones for 
the Gegenbauer and Hermite polynomials have been known for a long time 
\cite{Rain}. This happened because such formulae are not mentioned 
in any book devoted to special functions except for the cited textbook. 
Rainville derived them by making combined use of an appropriate
double power series and of the generating function of the orthogonal
polynomials under discussion. Owing to their importance for some
physical problems, we find it useful to give in this work 
a comprehensive overview based on an alternative approach. 
We prove that a property of the type \ (\ref{Legsum}) is shared 
by two one-parameter families of classical orthogonal polynomials, 
namely, the Gegenbauer and Laguerre polynomials. The main ingredients 
of our method are Cauchy's integral formula for the derivatives 
of a holomorphic function and specific functional equations satisfied 
by the generating functions of these well-known polynomials. 

The outline of the article is as follows. Section 2 deals with 
the Gegenbauer series, while in Sec. 3 the obtained formula is applied 
to the particular cases of the Legendre and Chebyshev polynomials. 
In Sec. 4  we review the Laguerre series. The sum of the series of Hermite polynomials is recovered in Sec. 5 in a straightforward manner, as well as 
an asymptotic limit of both the Gegenbauer and Laguerre series. In Sec. 6
we compute a similar power series of an arbitrary derivative of Mehler's 
generating function, finding a new formula. By way of illustration, 
Mehler's summation formula is employed in Sec. 7 to write in closed form 
the well-known propagator of a linear harmonic oscillator. 
Concluding remarks are presented in Sec. 8.
 
\vspace{4mm}

{\centering\section{GEGENBAUER POLYNOMIALS}}

\vspace{2mm}
\setcounter{equation}{0}
The generating function of the Gegenbauer polynomials of order $\lambda$, 
\begin{equation}
G_{0}^{(\lambda)}(t,w):=(1-2wt+t^2)^{-\lambda}, \qquad (\lambda \not=0,
\;\;\; \lambda>-\frac{1}{2}, \;\;\; w:=\cos\theta), \label{Gegfun}
\end{equation}
is holomorphic in the variable $t$ for $|t|<1$ and has the Taylor 
expansion \cite{HTFGeg}
\begin{equation}
G_{0}^{(\lambda)}(t,w)=\sum_{n=0}^{\infty}t^{n}C_{n}^{\lambda}(w), 
\qquad (w:=\cos\theta, \;\;\; |t|<1). \label{Geggen}
\end{equation}
Accordingly, Cauchy's integral formula for the derivatives of a 
holomorphic function (CIFD) \cite{Ahl, Hille} gives the following 
integral representation of the Gegenbauer polynomial $C_{l}^{\lambda}(w)$  
of degree $l$ and order $\lambda$:
\begin{equation}
C_{l}^{\lambda}(w)=\frac{1}{2{\pi}i}\int^{(0+)}dt \,\frac{1}{t^{l+1}}\,
G_{0}^{(\lambda)}(t,w). \label{GegCIF}  
\end{equation}
The contour of integration in eq.~(\ref{GegCIF}) is a simple loop
encircling the origin $t=0$ counterclockwise and situated entirely
within the open disk $|t|<1$. Note that the sum of the power series 
of interest, 
\begin{equation}
G_{l}^{(\lambda)}(t,w):=\sum_{n=l}^{\infty}\left(\begin{array}{c}n\\l
\end{array}\right)t^{n}C_{n}^{\lambda}(w), \qquad (w=:\cos\theta,
\;\;\; |t|<1, \;\;\; l=0,1,2,3,...), \label{Gegser}
\end{equation}
involves the $l$th-order  derivative of the generating function:
\begin{equation}
G_{l}^{(\lambda)}(t,w)=\frac{t^l}{l!}\left(\frac{\partial}{{\partial}t}
\right)^{l}G_{0}^{(\lambda)}(t,w). \label{Gegder}
\end{equation}
We employ CIFD again and find, after a translation of the variable
of integration,
\begin{equation}
G_{l}^{(\lambda)}(t,w)=\frac{t^l}{2{\pi}i}\int^{(0+)}ds \,\frac{1}{s^{l+1}}\,
G_{0}^{(\lambda)}(s+t,w). \label{CIFGs}
\end{equation}
We now take advantage of the following functional equation 
for the generating function \ (\ref{Gegfun}):
\begin{equation}
G_{0}^{(\lambda)}(s+t,w)=G_{0}^{(\lambda)}(t,w)\,G_{0}^{(\lambda)}(\xi, \eta),
\label{Gegfeq}
\end{equation}
where we have denoted
\begin{equation}
\xi:=\frac{s}{q}, \qquad \eta:=\frac{w-t}{q}. 
\label{Gegvar}
\end{equation}
Insertion of eq.~(\ref{Gegfeq}) into eq.~(\ref{CIFGs}), followed by 
the change of the integration variable to $\xi$ defined 
in eq.~(\ref{Gegvar}), yields
\begin{equation}
G_{l}^{(\lambda)}(t,w)=G_{0}^{(\lambda)}(t,w)\left(\frac{t}{q}\right)^l
\frac{1}{2{\pi}i}\int^{(0+)}d\xi \,\frac{1}{\xi^{l+1}}\,
G_{0}^{(\lambda)}(\xi, \eta). \label{Gegint}
\end{equation}
For real $t$, the obvious inequality ${\eta}^2 \leqq 1$ allows us to use 
eqs.~(\ref{Gegfun}) and \ (\ref{GegCIF}) and find the formula
\begin{eqnarray}
&&G_{l}^{(\lambda)}(t,w)=q^{-2\lambda}\left(\frac{t}{q}\right)^l
C_{l}^{\lambda}\left(\frac{w-t}{q}\right),
\nonumber \\ &&(\lambda \not=0, \;\;\; \lambda>-\frac{1}{2}, 
\;\;\; w:=\cos\theta, \;\;\; |t|<1, \;\;\;l=0,1,2,3,...). \label{Gegsum}
\end{eqnarray}
The validity of eq.~(\ref{Gegsum}) can be extended by analytic 
continuation to any complex value $t$ inside the unit circle.
An equivalent form of eq.~(\ref{Gegsum}), 
\begin{equation}
\sum_{m=0}^{\infty}\left(\begin{array}
{c}l+m\\l\end{array}\right)t^{m}C_{l+m}^{\lambda}(w)
=q^{-2\lambda-l}C_{l}^{\lambda}\left(\frac{w-t}{q}\right), \label{Gegdec}
\end{equation}
coincides with the formula written by Rainville \cite{RainGeg}.
 
\vspace{4mm}

{\centering\section{LEGENDRE AND CHEBYSHEV POLYNOMIALS}}

\vspace{2mm}
\setcounter{equation}{0}
Recall that the Legendre polynomials are the Gegenbauer polynomials 
of order $\lambda=\frac{1}{2}$:  
\begin{equation}
P_{l}(w)=C_{l}^{\frac{1}{2}}(w). \label{Legpol}  
\end{equation}
Therefore, by setting $\lambda=\frac{1}{2}$ in eqs.~(\ref{Gegser}) and
~(\ref{Gegsum}), we recover eqs.~(\ref{Legser}) and
~(\ref{Legsum}), respectively. For $\lambda=\frac{1}{2}$, eq.~(\ref{Gegdec})
reduces to the equivalent formula chosen by Rainville \cite{RainLeg}:
\begin{equation}
\sum_{m=0}^{\infty}\left(\begin{array}
{c}l+m\\l\end{array}\right)t^{m}P_{l+m}(w)=\left(\frac{1}{q}\right)^{1+l}
P_{l}\left(\frac{w-t}{q}\right). \label{RainLeg}
\end{equation}

The Chebyshev polynomials of the first kind,
\begin{equation}
T_{l}(w):=\cos(l\theta),  \qquad (w:=\cos\theta, \;\;\;  
l=0,1,2,3,...), \label{Che1}   
\end{equation}
are the limiting case $\lambda=0$ of the Gegenbauer polynomials, 
as follows \cite{HTFC1}:
\begin{equation}
T_{0}(w)=C_{0}^{0}(w), \label{Chezer}
\end{equation}
\begin{equation}
T_{l}(w)=\frac{l}{2}\,C_{l}^{0}(w), \qquad (l=1,2,3,...),  \label{Chenat}
\end{equation}
with \cite{HTFGz}
\begin{equation}
C_{0}^{0}(w):=1, \label{Gegzer}
\end{equation}
\begin{equation}
C_{l}^{0}(w):=\lim_{\lambda\rightarrow 0}\left[\frac{1}{\lambda}\,
C_{l}^{\lambda}(w)\right], \qquad (l=1,2,3,...).  \label{Gegnat}
\end{equation}
Hence the generating function of the polynomials \ (\ref{Gegnat})
is the derivative
\begin{equation}
G_{0}^{(0)}(t,w):=\frac{\partial G_{0}^{(\lambda)}(t,w)}
{\partial \lambda}\Bigg|_{\lambda=0}, \label{C1gfdef}
\end{equation}
while the functions studied here are given by the limit 
\begin{equation}
G_{l}^{(0)}(w):=\lim_{\lambda\rightarrow0}\left[\frac{1}{\lambda}\,
G_{l}^{(\lambda)}(t,w)\right], \qquad (l=1,2,3,...).  \label{C1fdef}
\end{equation}

The generating function \ (\ref{C1gfdef}), whose explicit form 
is the principal determination of the logarithm
\begin{equation}
G_{0}^{(0)}(t,w)=-\ln(1-2wt+t^2), \label{C1fun}
\end{equation}
is the sum of the power series \cite{HTFC1g}
\begin{equation}
G_{0}^{(0)}(t,w)=\sum_{n=1}^{\infty}t^{n}\frac{2}{n}\,T_{n}(w),
\qquad (w:=\cos\theta, \;\;\; |t|<1).  \label{C1gen}
\end{equation}
In turn, eq.~(\ref{Gegser}) leads us to write a function 
\ (\ref{C1fdef}) as the sum of the series
\begin{equation}
G_{l}^{(0)}(t,w)=\sum_{n=l}^{\infty}\left(\begin{array}{c}n\\l
\end{array}\right)t^{n}\frac{2}{n}\,T_{n}(w), \;\;\; (w:=\cos\theta,
\;\;\; |t|<1, \;\;\; l=1,2,3,...), \label{C1ser}
\end{equation}
while eq.~(\ref{Gegsum}) gives its explicit expression:
\begin{eqnarray}
G_{l}^{(0)}(t,w)=\left(\frac{t}{q}\right)^l
\frac{2}{l}\,T_{l}\left(\frac{w-t}{q}\right), \qquad (l=1,2,3,...). 
\label{C1sum}
\end{eqnarray}
In particular, for $l=1$ \cite{HTFC1s},
\begin{equation}
\sum_{n=0}^{\infty}t^{n}\,T_{n}(w)=\frac{1-wt}{1-2wt+t^2}. \label{C1ser1}
\end{equation}
Equations \ (\ref{C1ser}) and \ (\ref{C1sum}) also read:
\begin{equation}
\sum_{m=0}^{\infty}\left(\begin{array}{c}l+m\\l
\end{array}\right)t^{m}\frac{1}{l+m}\,T_{l+m}(w)=\left(\frac{1}{q}\right)^l
\frac{1}{l}\,T_{l}\left(\frac{w-t}{q}\right),
\qquad (l=1,2,3,...). \label{PTC1}
\end{equation}

The Chebyshev polynomials of the second kind,
\begin{equation}
U_{l}(w):=\frac{\sin[(l+1)\theta]}{\sin\theta},  \qquad
(w:=\cos\theta, \;\;\;  l=0,1,2,3,...), \label{Che2}   
\end{equation}
are the Gegenbauer polynomials of order $\lambda=1$ \cite{HTFC2}:
\begin{equation}
U_{l}(w)=C_{l}^{1}(w). \label{C2pol}  
\end{equation}
Accordingly, the sum of the series
\begin{equation}
G_{l}^{(1)}(t,w):=\sum_{n=l}^{\infty}\left(\begin{array}{c}n\\l
\end{array}\right)t^{n}U_{n}(w), \qquad (w:=\cos\theta,
\;\;\; |t|<1, \;\;\; l=0,1,2,3,...) \label{C2ser}
\end{equation}
is the special case $\lambda=1$ of eq.~(\ref{Gegsum}): 
\begin{eqnarray}
G_{l}^{(1)}(t,w)=q^{-2}\left(\frac{t}{q}\right)^l
U_{l}\left(\frac{w-t}{q}\right), \qquad (l=0,1,2,3,...). 
\label{C2sum}
\end{eqnarray}
An equivalent formula is given by eq.~(\ref{Gegdec}) written 
for $\lambda=1$:
\begin{eqnarray}
\sum_{m=0}^{\infty}\left(\begin{array}
{c}l+m\\l\end{array}\right)t^{m}U_{l+m}(w)=\left(\frac{1}{q}\right)^{2+l}
U_{l}\left(\frac{w-t}{q}\right). \label{PTC2}
\end{eqnarray}
 
\vspace{4mm}

{\centering\section{LAGUERRE POLYNOMIALS}}

\vspace{2mm}
\setcounter{equation}{0}
For the sake of completeness, we prove the similar property of the 
Laguerre polynomials, repeating step by step the argument from Section 2. 
The generating function of the Laguerre polynomials of order $\alpha$, 
\begin{equation}
S_{0}^{(\alpha)}(t,u):=(1-t)^{-\alpha-1}\exp\left(-\frac{ut}{1-t}\right),
\qquad (\alpha>-1), 
\label{Lagfun}
\end{equation}
is holomorphic in the variable $t$ inside the circle $|t|=1$ 
and has the Taylor expansion \cite{HTFLag}
\begin{equation}
S_{0}^{(\alpha)}(t,u)=\sum_{n=0}^{\infty}t^{n}L_{n}^{\alpha}(u), \qquad (|t|<1). \label{Laggen}
\end{equation}
Making use of CIFD, we write the Laguerre polynomial 
$L_{l}^{\alpha}(u)$ of degree $l$ and order $\alpha$ as an integral 
representation  \cite{corr}:
\begin{equation}
L_{l}^{\alpha}(u)=\frac{1}{2{\pi}i}\int^{(0+)}dt \, \frac{1}{t^{l+1}}
\, S_{0}^{(\alpha)}(t,u). \label{LagCIF}  
\end{equation}
Just as in eq.~(\ref{GegCIF}), the contour of integration encircles 
the origin $t=0$ once, counterclockwise, inside the unit circle. Now 
the power series under investigation,
\begin{equation}
S_{l}^{(\alpha)}(t,u):=\sum_{n=l}^{\infty}\left(\begin{array}{c}n\\l
\end{array}\right)t^{n}L_{n}^{\alpha}(u), \qquad (|t|<1, \;\;\; 
l=0,1,2,3,...), \label{Lagser}
\end{equation}
has the sum
\begin{equation}
S_{l}^{(\alpha)}(t,u)=\frac{t^l}{l!}\left(\frac{\partial}{{\partial}t}
\right)^{l}S_{0}^{(\alpha)}(t,u). \label{Lagder}
\end{equation}
Applying once again CIFD, we get with a translation of the variable
of integration: 
\begin{equation}
S_{l}^{(\alpha)}(t,u)=\frac{t^l}{2{\pi}i}\int^{(0+)}ds\, \frac{1}{s^{l+1}}\,
S_{0}^{(\alpha)}(s+t,u). \label{CIFLs}
\end{equation}
The generating function \ (\ref{Lagfun}) satisfies the functional equation 
\begin{equation}
S_{0}^{(\alpha)}(s+t,u)=S_{0}^{(\alpha)}(t,u)\,S_{0}^{(\alpha)}
\left(\frac{s}{1-t}\, , \, \frac{u}{1-t}\right). \label{Lagfeq}
\end{equation}
Combining it with the change of variable  
\begin{equation}
\zeta:=\frac{s}{1-t} \label{Lagvar}
\end{equation}
in the integral, eq.~(\ref{CIFLs}) becomes
\begin{equation}
S_{l}^{(\alpha)}(t,u)=S_{0}^{(\alpha)}(t,u)\left(\frac{t}{1-t}\right)^l
\frac{1}{2{\pi}i}\int^{(0+)}d\zeta \,\frac{1}{\zeta^{l+1}}\,
S_{0}^{(\alpha)}\left(\zeta,\, \frac{u}{1-t}\right). \label{Lagint}
\end{equation}
On account of eqs.~(\ref{Lagfun}) and \ (\ref{LagCIF}), 
eq.~(\ref{Lagint}) finally yields the sum of the series 
\ (\ref{Lagser}): 
\begin{eqnarray}
&&S_{l}^{(\alpha)}(t,u)=(1-t)^{-\alpha-1}\exp\left(-\frac{ut}{1-t}\right)
\left(\frac{t}{1-t}\right)^{l}L_{l}^{\alpha}\left(\frac{u}{1-t}\right),
\nonumber \\ &&(\alpha>-1, \;\;\; |t|<1, \;\;\; l=0,1,2,3,...). 
\label{Lagsum}
\end{eqnarray}
Note that eqs.~(\ref{Lagser}) and \ (\ref{Lagsum}) are equivalent 
to the formula preferred by Rainville \cite{RainLag}:
\begin{equation}
\sum_{m=0}^{\infty}\left(\begin{array}
{c}l+m\\l\end{array}\right)t^{m}L_{l+m}^{\alpha}(u)=(1-t)^{-\alpha-1}
\exp\left(-\frac{ut}{1-t}\right)\left(\frac{1}{1-t}\right)^l
L_{l}^{\alpha}\left(\frac{u}{1-t}\right). \label{RainLag}
\end{equation}
 
\vspace{4mm}

{\centering\section{HERMITE POLYNOMIALS}}

\vspace{2mm}
\setcounter{equation}{0}
The generating function of the Hermite polynomials is the exponential
\begin{equation}
W_{0}(z,x):=\exp(2xz-z^2).  \label{Herfun}
\end{equation}
Being an entire function, the Taylor series \cite{HTFHer}
\begin{equation}
W_{0}(z,x)=\sum_{n=0}^{\infty}\frac{z^{n}}{n!}\,H_{n}(x) \label{Hergen}
\end{equation}
converges in the whole complex $z$-plane. The more general series 
we are interested in,
\begin{equation}
W_{l}(z,x):=\sum_{n=l}^{\infty}\left(\begin{array}{c}n\\l
\end{array}\right)\frac{z^{n}}{n!}\,H_{n}(x), \qquad (l=0,1,2,3,...), 
\label{Herser}
\end{equation}
has the sum
\begin{equation}
W_{l}(z,x))=\frac{z^l}{l!}\left(\frac{\partial}{{\partial}z}
\right)^{l}W_{0}(z,x). \label{Herder}
\end{equation}
In order to evaluate it, we apply the method of Secs. 2 and 3, 
taking advantage of the functional equation satisfied 
by the generating function ~(\ref{Herfun}):
\begin{equation}
W_{0}(v+z,x)=W_{0}(z,x)\,W_{0}(v,x-z). \label{Herfeq}
\end{equation}
The result is
\begin{equation}
W_{l}(z,x)=\exp(2xz-z^2)\,\frac{t^{l}}{l!}\,H_{l}(x-z),
\qquad (l=0,1,2,3,...), \label{Hersum}
\end{equation}
and may be written in the equivalent form chosen by Rainville \cite{RainHer}:
\begin{equation}
\sum_{m=0}^{\infty}\frac{z^{m}}{m!}\,H_{l+m}(x)=\exp(2xz-z^2)\,H_{l}(x-z). 
\label{Herdec}
\end{equation}

However, eq.~(\ref{Hersum}) can be established as an asymptotic case
of the similar formula either for the Gegenbauer polynomials, 
eq.~(\ref{Gegsum}), or for the Laguerre polynomials, eq.~(\ref{Lagsum}).

In the first case, our starting point is the limit \cite{AARGeg}
\begin{equation}
W_{0}(z,x)=\lim_{\lambda\rightarrow\infty}\left[G_{0}^{(\lambda)}
({\lambda}^{-\frac{1}{2}}z, {\lambda}^{-\frac{1}{2}}x)\right]. 
\label{Ggenas}
\end{equation}
Hence
\begin{equation}
\frac{1}{l!}H_{l}(x)=\lim_{\lambda\rightarrow\infty}\left[{\lambda}^
{-\frac{l}{2}}C_{l}^{\lambda}({\lambda}^{-\frac{1}{2}}x)\right]. 
\label{Gpolas}
\end{equation}
Making use of eqs. \ (\ref{Herder}) and \ (\ref{Gegder}), we get
\begin{equation}
W_{l}(z,x)=\lim_{\lambda\rightarrow\infty}\left[G_{l}^{(\lambda)}
({\lambda}^{-\frac{1}{2}}z, {\lambda}^{-\frac{1}{2}}x)\right]. 
\label{Gsumas}
\end{equation}
Substitution of eqs. \ (\ref{Gegsum}), \ (\ref{Ggenas}), and 
\ (\ref{Gpolas}) into eq.~(\ref{Gsumas}) yields again the result
\ (\ref{Hersum}).

In the second case, we start from the identity \cite{AARLag}
\begin{equation}
W_{0}(z,x)=\lim_{\alpha\rightarrow\infty}\left\{S_{0}^{(\alpha)}
\left(\left(\frac{2}{\alpha}\right)^{\frac{1}{2}}z,\, 
{\alpha}\left[1-\left(\frac{2}{\alpha}\right)^{\frac{1}{2}}x\right]
\right)\right\}, \label{Lgenas}
\end{equation}
which gives the limit
\begin{equation}
\frac{1}{l!}H_{l}(x)=\lim_{\alpha\rightarrow\infty}\left\{\left(
\frac{2}{\alpha}\right)^{\frac{l}{2}}L_{n}^{\alpha}\left({\alpha}
\left[1-\left(\frac{2}{\alpha}\right)^{\frac{1}{2}}x\right]
\right)\right\}. \label{Lpolas}
\end{equation}
On the other hand, taking note of eqs.~(\ref{Herder}) and 
~(\ref{Lagder}), we find:
\begin{equation}
W_{l}(z,x)=\lim_{\alpha\rightarrow\infty}\left\{S_{l}^{(\alpha)}
\left(\left(\frac{2}{\alpha}\right)^{\frac{1}{2}}z,\, 
{\alpha}\left[1-\left(\frac{2}{\alpha}\right)^{\frac{1}{2}}x\right]
\right)\right\}. \label{Lsumas}
\end{equation}
By inserting eq.~(\ref{Lagsum}) as well as the limits ~(\ref{Lgenas}) 
and ~(\ref{Lpolas}) into eq.~(\ref{Lsumas}), we recover once more 
eq.~(\ref{Hersum}). 
 
\vspace{4mm}

{\centering\section{A GENERALIZATION OF MEHLER'S SUMMATION FORMULA}}

\vspace{2mm}
\setcounter{equation}{0}
We intend to evaluate the sum of the power series
\begin{equation}
M_{l}(z,x,y):=\sum_{n=l}^{\infty}\left(\begin{array}{c}n\\l
\end{array}\right)\frac{z^{n}}{n!\,2^n}\,H_{n}(x)H_{n}(y), \qquad
(|z|<1, \;\;\; l=0,1,2,3,...). \label{Mehser}
\end{equation}
Its special case $l=0$ is Mehler's power series expansion,
\begin{equation}
M_{0}(z,x,y)=\sum_{n=0}^{\infty}\frac{z^{n}}{n!\,2^n}\,H_{n}(x)H_{n}(y), 
\qquad (|z|<1), \label{Mehgen}
\end{equation}
of the generating function \cite{HTFMeh,HBMeh,AARMeh}
\begin{equation}
M_{0}(z,x,y):=\left(1-z^2\right)^{-\frac{1}{2}}\exp\left\{\frac{1}{1-z^2}
\left[2xyz-\left(x^2+y^2\right)z^2\right]\right\}. \label{Mehfun}
\end{equation}
The analytic function\ (\ref{Mehfun}) factors in the following way 
\cite{HBMfac}:
\begin{equation}
M_{0}(z,x,y)=
S_{0}^{\left(-\frac{1}{2}\right)}\left(z,\frac{1}{2}(x-y)^2\right)
S_{0}^{\left(-\frac{1}{2}\right)}\left(-z,\frac{1}{2}(x+y)^2\right), 
\qquad (|z|<1), \label{Mehfac}
\end{equation}
where $S_{0}^{\left(-\frac{1}{2}\right)}(t,u)$ is the generating function 
\ (\ref{Lagfun}) of the Laguerre polynomials of order 
$\alpha=-\frac{1}{2}$.
On the other hand, the function\ (\ref{Mehser}) involves a $l$th-order 
partial derivative of Mehler's generating function \ (\ref{Mehfun}):
\begin{equation}
M_{l}(z,x,y))=\frac{z^l}{l!}\left(\frac{\partial}{{\partial}z}
\right)^{l}M_{0}(z,x,y). \label{Mehder}
\end{equation}
Insertion of the factorization\ (\ref{Mehfac}) into 
eq.~(\ref{Mehder}), followed by application of Leibniz' rule
and use of eq.~(\ref{Lagder}) with $\alpha=-\frac{1}{2}$, leads us
to a finite expansion:
\begin{eqnarray}
&&M_{l}(z,x,y)=\sum_{m=0}^{l}
S_{l-m}^{\left(-\frac{1}{2}\right)}\left(z,\frac{1}{2}(x-y)^2\right)
S_{m}^{\left(-\frac{1}{2}\right)}\left(-z,\frac{1}{2}(x+y)^2\right), 
\nonumber \\&&(\,|z|<1, \quad l=0,1,2,3,...). \label{Mehsum}
\end{eqnarray}
By substituting the series sum~(\ref{Lagsum}) and employing
again eq.~(\ref{Mehfac}), we establish the following formula:
\begin{eqnarray}
&&M_{l}(z,x,y)=\left(\frac{z}{1-z}\right)^{l}M_{0}(z,x,y)\times
\nonumber \\&&\sum_{m=0}^{l}\left(-\frac{1-z}{1+z}\right)^{m}
L_{l-m}^{-\frac{1}{2}}\left(\frac{(x-y)^2}{2(1-z)}\right)
L_{m}^{-\frac{1}{2}}\left(\frac{(x+y)^2}{2(1+z)}\right),
\nonumber \\&&(\,|z|<1, \quad l=0,1,2,3,...). \label{MehsL}
\end{eqnarray}
The function \ (\ref{MehsL}) is a finite sum involving Laguerre 
polynomials of order $\alpha=-\frac{1}{2}$. It can equally be expressed 
in terms of Hermite polynomials, via the identity \cite{HTFLH}
\begin{equation}
H_{2m}(x)=(-1)^{m}m!\,2^{2m}L_{m}^{-\frac{1}{2}}(x^2).
\label{LagHer}
\end{equation}
We get thus an alternative explicit formula:
\begin{eqnarray}
&&M_{l}(z,x,y)=\frac{(-1)^l}{l!\,2^{2l}}\left(\frac{z}{1-z}\right)^{l}
M_{0}(z,x,y)\times
\nonumber \\&&\sum_{m=0}^{l}\left(\begin{array}{c}l\\m
\end{array}\right)\left(-\frac{1-z}{1+z}\right)^{m}
H_{2(l-m)}\left(\frac{x-y}{[2(1-z)]^{\frac{1}{2}}}\right)
H_{2m}\left(\frac{x+y}{[2(1+z)]^{\frac{1}{2}}}\right),
\nonumber \\&&(\,|z|<1, \quad l=0,1,2,3,...). \label{MehsH}
\end{eqnarray}
 
\vspace{4mm}

{\centering\section{PROPAGATOR OF A LINEAR HARMONIC OSCILLATOR}}

\vspace{2mm}
\setcounter{equation}{0}
In nonrelativistic quantum mechanics it is instructive to evaluate 
the propagator of a linear harmonic oscillator by employing 
Mehler's summation formula, eqs.~(\ref{Mehgen})-~(\ref{Mehfun}).  
In the coordinate representation, the propagator  is defined 
as the probability amplitude for finding the particle at some point 
$x_b$, at time $t_b$, when it has originally been at another point $x_a$, 
at time $t_a$:
\begin{equation}
K(x_b, t_b ; x_a, t_a):=\langle x_b, t_b\,|\, x_a, t_a\rangle. 
\label{propag}
\end{equation}
For an oscillator of mass $m$ and classical angular frequency $\omega$ 
it is convenient to denote 
$\alpha:=\left(\frac{m\omega}{\hbar}\right)^{\frac{1}{2}}>0$. 
We recall the eigenvalues 
\begin{equation}
E_n=\left(n+\frac{1}{2}\right)\hbar\omega, \qquad (n=0,1,2,3,...),
\label{levels}
\end{equation}
as well as the corresponding eigenfunctions of its energy operator $\hat H$,
\begin{equation}
u_n(x)={\left(\frac{\alpha}{\pi^{1/2}n!\,2^n}\right)}^{1/2}
\exp\left(-\frac{1}{2}\alpha^2x^2\right)\,H_n(\alpha x), 
\qquad (n=0,1,2,3,...).
\label{eigenf}
\end{equation}
Making use of the time-evolution operator in the Schr\"odinger picture,
\begin{equation}
\hat U(t_b,\, t_a)=\exp\left[-\frac{i}{\hbar}(t_b-t_a)\hat H\right],
\label{evolution}
\end{equation}
the transition probability amplitude \ (\ref{propag}) reads:
\begin{equation}
\langle x_b, t_b\,|\, x_a, t_a\rangle=\langle x_b \,|\hat U(t_b,\, t_a)|
\, x_a\rangle. 
\label{time}
\end{equation}
Insertion of  eq.~(\ref{evolution}) into eq.~(\ref{time}) yields, 
via the spectral decomposition of the energy $\hat H$, 
the standard eigenfunction expansion of the propagator:
\begin{equation}
K(x_b, t_b ; x_a, t_a)=\sum_{n=0}^{\infty}
\exp\left[-i\,\frac{E_n}{\hbar}(t_b-t_a)\right]u_n(x_b)\,u_n^*(x_a). 
\label{expansion}
\end{equation}
Substitution of the energy levels ~(\ref{levels}) and eigenfunctions
~(\ref{eigenf}) into eq.~(\ref{expansion}) enables us to exploit 
Mehler's expansion ~(\ref{Mehgen}) to get a compact formula:
\begin{eqnarray}
&&K(x_b, t_b ; x_a, t_a)=\frac{\alpha}{\pi^{1/2}}
\exp\left[-i\,\frac{\omega}{2}(t_b-t_a)-\frac{1}{2}\,\alpha^2\left(x_b^2
+x_a^2\right)\right]\times
\nonumber \\ && M_0\left(\exp\left[-i\omega(t_b-t_a)\right], 
\alpha x_b,\, \alpha x_a\right). \label{compact}
\end{eqnarray}
Equations ~(\ref{Mehfun}) and ~(\ref{compact}) finally give the exact 
closed-form expression of the propagator of the one-dimensional 
harmonic oscillator \cite{Merz}:
\begin{eqnarray}
&&K(x_b, t_b ; x_a, t_a)=\left\{\frac{-im\omega}{2\pi\hbar
\sin\left[\omega(t_b-t_a)\right]}\right\}^{\frac{1}{2}}\times
\nonumber \\ &&\exp\left(\frac{im\omega}
{2\hbar\sin\left[\omega(t_b-t_a)\right]}
\left\{\left(x_b^2+x_a^2\right)\cos\left[\omega(t_b-t_a)\right]
-2x_b\,x_a\right\}\right). \label{explicit}
\end{eqnarray}
Since Mehler's summation of the eigenfunction series ~(\ref{expansion})
does not always fulfil the convergence requirement from eq.~(\ref{Mehgen}),
the propagator ~(\ref{explicit}) is a distribution involving $\delta$-type 
singularities when the time difference $t_b-t_a$ is a multiple 
of the classical half-period of oscillation ${\pi}/{\omega}$. 
For instance, eq.~(\ref{time}), and eq.~(\ref{expansion}) as well, 
displays the equal-time values 
\begin{equation}
K(x_b, t ; x_a, t)=\delta(x_b-x_a).  \label{delta}
\end{equation}

We stress that the transition probability amplitudes 
of the type ~(\ref{propag}) are at heart of Feynman's path-integral 
formulation of quantum mechanics. In particular, for the linear harmonic 
oscillator the propagator ~(\ref{propag}) has been evaluated exactly 
in this framework \cite{Feynman, Klein}. However, the explicit 
formula ~(\ref{explicit}) can also be obtained in an alternative 
simpler way \cite{Merz}. Had one started from eq.~(\ref{explicit}), 
one could then find, conversely, the energy levels ~(\ref{levels}) 
and eigenfunctions ~(\ref{eigenf}). Indeed, insertion of Mehler's 
power series ~(\ref{Mehgen}) into eq.~(\ref{compact}) allows one 
to compare the resulting formula with the general eigenfunction 
expansion ~(\ref{expansion}). One thus readily gets the well-known 
solution of this energy eigenvalue problem. 
  
\vspace{4mm}

{\centering\section{CONCLUSIONS}}

\vspace{2mm}
\setcounter{equation}{0}
The sum of a series of the type ~(\ref{Legser}) involves the $l$th-order 
derivative of the generating function of the orthogonal polynomials. 
We have evaluated it in a straightforward way for both the Gegenbauer and 
Laguerre polynomials taking advantage of the analytic structure 
of their generating functions. The explicit sum is found to be a product
of three factors: the generating function, the $l$th power of a specific 
variable, and the corresponding orthogonal polynomial of degree $l$ 
in another specific variable. However, the limiting case of the Chebyshev polynomials of the first kind stands out since their generating function 
is missing in the explicit formula. Note that the above-mentioned result 
is by no means general: it holds neither for arbitrary Jacobi polynomials, 
nor for some nonclassical orthogonal polynomials \cite {AARpol}. This is 
due to the analytic peculiarities of the generating functions in question.
Along the same lines, we have succeeded in evaluating the higher-order 
derivatives of Mehler's generating function as finite sums involving 
products of pairs of Hermite polynomials of even degrees, eq.~(\ref{MehsH}).
The importance of Mehler's formula in quantum mechanics is finally
emphasized.

\vspace{4mm}

{\centering\subsubsection*{Acknowledgements}}

\vspace{2mm}

This work was supported by the Romanian Ministry of Education and Research
through Grant IDEI-995/2007 for the University of Bucharest.

\vspace{3mm}

\begin{center}

\end{center}

\begin{thebibliography}{25}
\bibitem{HTFC1g} Ref. \ cite{HTFGeg}, Vol. 2, p. 186, eq. (30):
there is a mistake that can be corrected by omitting unity 
in the l.h.s. of this equation.
\bibitem{corr} There is a mistake in eq.(B3) of Ref. \cite{PTB}. 
The correct formula reads
\begin{eqnarray*}
L_{n}^{(\alpha)}(u)=\frac{1}{2{\pi}i}\int^{(0+)}dt\, \frac{1}{t^{n+1}}
\, T_{0}^{(\alpha)}(t,u)
\end{eqnarray*}
and coincides with eq.~(\ref{LagCIF}) of the present work.
\end{thebibliography}
\end{document}